# Investigation of La and Al substitution on the spontaneous polarization and lattice dynamics of the $Pb_{(1-x)}La_xTi_{(1-x)}Al_xO_3$ ceramics

Arun Kumar Yadav[1], Anita Verma[1], Sunil Kumar[1], Velaga Srihari[2], A. K. Sinha[3,4], V. Raghavendra Reddy[5], Shun Wei Liu[6], Sajal Biring[6*], Somaditya Sen[1*]

[1]Discipline of Metallurgy Engineering and Materials Science, Indian Institute of Technology Indore, Khandwa Road, Indore-453552, India
[2]High Pressure & Synchrotron Radiation Physics Division, Bhabha Atomic Research Centre, 400085, Mumbai, India
[3]HXAL, SUS, Raja Ramanna Centre for Advanced Technology, Indore-452013, India
[4]Homi Bhabha National Institute, Anushaktinagar, Mumbai-400098, India
[5]UGC-DAE Consortium for Scientific Research, University Campus, Khandwa Road, Indore-452001, India
[6]Electronic Engg., Ming Chi University of Technology, New Taipei City, Taiwan

**Abstract:** The phase purity and crystal structure of $Pb_{(1-x)}La_xTi_{(1-x)}Al_xO_3$ ($0 \leq x \leq 0.25$) samples (synthesized via sol-gel process) were confirmed using synchrotron x-ray powder diffraction (XRD) (wavelength, $\lambda = 0.44573$ Å). Rietveld analyses of powder x-ray diffraction data confirmed the tetragonal structure for compositions with $x \leq 0.18$ and cubic structure for the sample with $x=0.25$. Temperature-dependent XRD was performed to investigate the structural change from tetragonal to cubic structure phase transition. Raman spectroscopy at room temperature also confirmed this phase transition with composition. Field emission scanning electron microscopy (FESEM) provided information about surface morphology while an energy dispersive x-ray spectrometer (EDS) attached with FESEM confirmed the chemical compositions of samples. Temperature and frequency dependent dielectric studies showed that the tetragonal to cubic phase transition decreased from 680 K to 175 K with increase in the $x$ from 0.03 to 0.25, respectively. This is correlated with the structural studies. Electric field dependent spontaneous polarization showed proper ferroelectric loop for $0.06 \leq x \leq 0.18$ belonging to a tetragonal phase while $x \geq 0.25$ the spontaneous polarization vanishes. Bipolar strain versus electric field revealed a butterfly loop for $0.06 \leq x \leq 0.18$ compositions. Energy storage efficiency initially increases nominally with substitution but beyond $x=0.18$ enhances considerably.

## Introduction

Ferroelectric/piezoelectric materials are recognized as excellent sensors and actuators due to their cost-effective and design flexibility with excellent functionalities[1, 2]. Industrial applications of piezoelectric are found in ultrasonic nondestructive devices, fuel injection valves, piezoelectric motors, etc. Hence, they are essential components to civil, aerospace, mining, automotive industries etc. [3] Usage of these materials have increased exponentially in the last decade. With switchable spontaneous polarization, ferroelectrics are the soul of piezoelectric transducers and sensors. Ferroelectrics are also the core parts of many energy conversion devices, such as ultrasonic medical diagnostic apparatus, ultrasonic nondestructive detectors, pyroelectric infrared sensors and magnetoelectric sensors. Amongst the various ferroelectrics, perovskites are technologically the most prominent due to their strong polarization, good response and multiple phases [4]. Perovskite oxides, with chemical formula $ABO_3$, are a wide range of compounds, with a larger twelve-coordinated $A$ cation and a smaller six-coordinated $B$ cation. The structure is commonly represented by a cuboid unit cell with $A$ cations at the corners, $B$ cations at the centre while the $O$ anions are located around the central point of each face. Lattice distortions in perovskite crystals arise out of off-centred ionic displacements and order-disorder transitions which generate reversible spontaneous polarization.[5]

Lead titanate ($PbTiO_3$) is a classic example of a displacive type ferroelectric material having highest tetragonal distortion $\sim c/a=1.06$ creating a strained lattice which makes it an excellent piezoelectric, dielectric and pyroelectric material at room temperature.[6, 7] $PbTiO_3$ has a structural phase transition from tetragonal ($P4mm$) to cubic ($Pm3m$) at Curie temperature, $T_c \sim$ 490°C. The non-centrosymmetry is a requirement for its ferroelectric functionalities and high spontaneous polarization. $PbTiO_3$ based compounds find pronounced applications as sensors, actuators, capacitors, nonvolatile memories, ultrasonic transducers, strong piezoelectric, electromechanical and optoelectronic materials, etc. [8]

Theoretical first principle calculations link *Ti-O* and *Pb-O* hybridizations to ferroelectricity in $PbTiO_3$ perovskite.[9, 10] $Pb\ 6s^2$ lone pair helps in stabilizing the large tetragonal strain ($\sim 6\%$). However, $Pb$ is not environmental friendly because of its highly toxic nature. So, $Pb$ free or reduced $Pb$ solid solutions with good ferroelectric properties are under extensive investigation. Based on this considerations, several $Pb$ reduced ferroelectric materials have been investigated, such as $Pb(Mg_{1/3}Nb_{2/3})O_3$-$PbTiO_3$, $Ba(Mg_{1/3}Nb_{2/3})O_3$−$PbTiO_3$, $Ba(Zn_{1/3}Nb_{2/3})O_3$−$PbTiO_3$, $Ba(Yb_{1/2}Nb_{1/2})O_3$−$PbTiO_3$, $Ba(Sc_{1/2}Nb_{1/2})O_3$−$PbTiO_3$, $BaSnO_3$−$PbTiO_3$, $(1–x)PbTiO_3$–$xBi(Ni_{1/2}Ti_{1/2})O_3$, $Pb(In_{1/2}$-$Nb_{1/2})O_3$–$Pb(Mg_{1/3}Nb_{2/3})O_3$–$PbTiO_3$, $PbHfO_3$–$Pb(Mg_{1/3}Nb_{2/3})O_3$–$PbTiO_3$, $Pb(Sn,Ti)O_3$–$Pb(Mg_{1/3}Nb_{2/3})O_3$–$xPbTiO_3$, $Pb(Ho_{1/2}Nb_{1/2})O_3$–$Pb(Mg_{1/3}Nb_{2/3})O_3$–$PbTiO_3$, $Pb(Y_{1/2}Nb_{1/2})O_3$–$Pb(Mg_{1/3}Nb_{2/3})O_3$–$PbTiO_3$ and $(1-x)PbTiO_3$–$xBiFeO_3$ $PbTiO_3$–$Bi(Zn_{1/2}Ti_{1/2})O_3$ and $PbTiO_3$–$Bi(Ni_{1/2}Ti_{1/2})O_3$ and $PbTiO_3$–$BiFeO_3$ etc. [11-17] $Pb$ based compounds show a diffuse type of phase transition behaviour. In dielectric response, diffuse type behaviour has a broad peak with temperature rather than sharp peak in normal ferroelectrics. The diffuse nature dielectric peak in compounds with mixed cations at lattice sites is due to the variation of phase transition temperature in different localized micro-regions.[18, 19]

Dielectric and ferroelectric properties of pure $PbTiO_3$ phase has been reported in many reports to be a difficult investigation due to limitations of preparation of phase pure, sufficiently dense and tough $PbTiO_3$ pellets.[7, 20-23] Large volume changes upon cooling through the cubic-tetragonal phase transformation, at $T_m \sim 490°C$, lead to internal stresses in the pure $PbTiO_3$ pellets. This generates fragile pellets which are difficult to handle during characterization.

Modified PbTiO$_3$, have been attempted by adding additives, that results in degradation of ferroelectric properties.

The motivation of the present work is to explore the effects of newly synthesized Pb$_{(1-x)}$La$_x$Ti$_{(1-x)}$Al$_x$O$_3$ (0≤$x$≤0.25) compounds have been investigated with structural, vibrational, dielectric, and ferroelectric properties. Due to high non-centrosymmetric tetragonal lattice strain ($c/a$~1.064) in pure PbTiO$_3$, it is not possible to use it in ceramics form in ferroelectric and piezoelectric applications. La/Al substituted PbTiO$_3$ samples were easily processable in ceramic form without sacrificing the ferroelectric functionalities.

**Synthesis and Experimental**

Pb$_{(1-x)}$La$_x$Ti$_{(1-x)}$Al$_x$O$_3$ (0≤$x$≤0.25) (PLTA) ceramics were prepared using sol-gel process. Precursors used to synthesize these materials were lead (II) nitrate, lanthanum nitrate, aluminium nitrate and dihydroxy bis (ammonium lactate) titanium (IV) (50% w/w aqua solution). All the precursors were Puratronic grade (purity >99.999 from Alfa Aesar). The precursors were selected on the basis of solubility in doubly deionized (DI) water. Stoichiometric solutions of each precursor were prepared with DI water in separate glass beakers. Lanthanum solution was added to the titanium solution. Aluminium and lead solutions were added to mixture subsequently. 5% excess lead nitrate was used accounting for the volatile nature of *Pb* during sintering at high temperatures. The mixed solution was stirred at normal temperature for 30 minutes to ensure proper homogeneity in the composition. A gel former was prepared in a separate beaker by using a solution of citric acid and ethylene glycol in 1:1 molar ratio. The homogeneous solution containing the requisite ions were poured in the gel former and vigorous stirring and heating (~85°C) followed on the hot plate continuously until a gel was formed. Gels were burnt in the open air inside a fume hood. Burnt powders were ground and thereafter heated at 500°C for 12h to get rid of trapped polymers and nitrates. The resultant powders were grounded with the mortar pestle and re-heated at 700°C for 12h. After that, the powders were mixed with 5% PVA solution (a binder) and pressed into pellet form of 10 mm diameter and 1.5 mm thickness using a uniaxial press. These pellets were sintered first at 600°C for 6h (to burn out the binder) and continuous 1150°C for 6h. Note that we had covered the pellets with powders of the same composition to reduce *Pb* loss. Mechanically strong pellets were formed at this optimum temperature. However, it has been reported that *Pb* evaporates from the pellet surface layers during sintering upto 200–250 nm deep, we have removed a similar amount from the pellets before performing any characterization. One of the pellets of each composition was ground for structural characterization using a XRD [Bruker D2 Phaser]. We have done synchrotron-based XRD experiments to confirm the purity and structure of the phase further. Synchrotron-based powder angle dispersive x-ray diffraction measurements (ADXRD) was carried out at the Extreme Conditions Angle Dispersive/Energy dispersive x-ray diffraction (EC-AD/ED-XRD) beamline (BL-11) at Indus-2 synchrotron source, Raja Ramanna Centre for Advanced Technology (RRCAT), India. Measurements were carried out in capillary mode. The crushed powders were filled in the capillary of 0.2 mm diameters. The diameter of capillary tube was selected taking into account of linear absorption co-efficient of the sample and 25% packing density of material in capillary. The capillary was rotated at ~150 rpm to reduce the orientation effects. From the white light of the bending magnet, a wavelength of λ= 0.44573Å was selected using a Si(111) channel-cut monochromator and then focused with a Kirkpatrick-Baez mirror or K-B mirror for ADXRD measurements. A MAR345 image plate detector (which is an area detector) was used to collect two-dimensional diffraction data. The sample to detector distance and the wavelength of the beam were calibrated using NIST standards $LaB_6$ and $CeO_2$. Calibration and conversion/integration of *2D* diffraction data to *1D*, intensity versus 2theta, was carried out using FIT2D software [24]. High-temperature structural study for *x*=0.09 composition was performed with synchrotron-based source (wavelength ~ 0.80165 Å) at BL-12, Indus-2 Beamline in RRCAT Indore, India.

The HR Raman spectroscopy was done with Czerny-Turner type achromatic spectrograph with the spectral resolution of 0.4 $cm^{-1}$/pixel. The source of wavelength excitation is 632.8 nm. Microstructure and grain size of the sintered pellets were investigated by Supra55 Zeiss field emission scanning electron microscope. Electrodes were painted on both sides of the sintered pellets using high-temperature silver paste. The silver coated pellets on either side were cured at 550°C for 10 minutes to improve the pellet electrode adhesion. Before doing the measurement, we heated the samples at 200°C for 10 minutes to get rid of any adhered surface moisture. A dielectric response was measured using a Newton's 4[th] Ltd PSM 1735 phase sensitive LCR meter (with the signal strength of ~1V). Ferroelectric (P-E) loop measurements were carried on the pellets immersed in silicone oil (to prevent electric arcing at high voltages) using a ferroelectric loop (P-E) tracer of M/s Radiant Instruments, USA.

**Results and discussion**

XRD patterns [Fig.1a] reveal that PLTA crystallographic structure resembles pure $PbTiO_3$ structure. The x-ray diffraction peaks are merging with increase in composition. The peak intensity of (001) and (110) reduced followed by a merger with (100) and (101) respectively as shown in Fig. 1d. Other peaks also related to a tetragonal phase merged to cubic structure too. However, beyond *x*=0.18 there was no splitting in peaks, and a near cubic phase was evident.

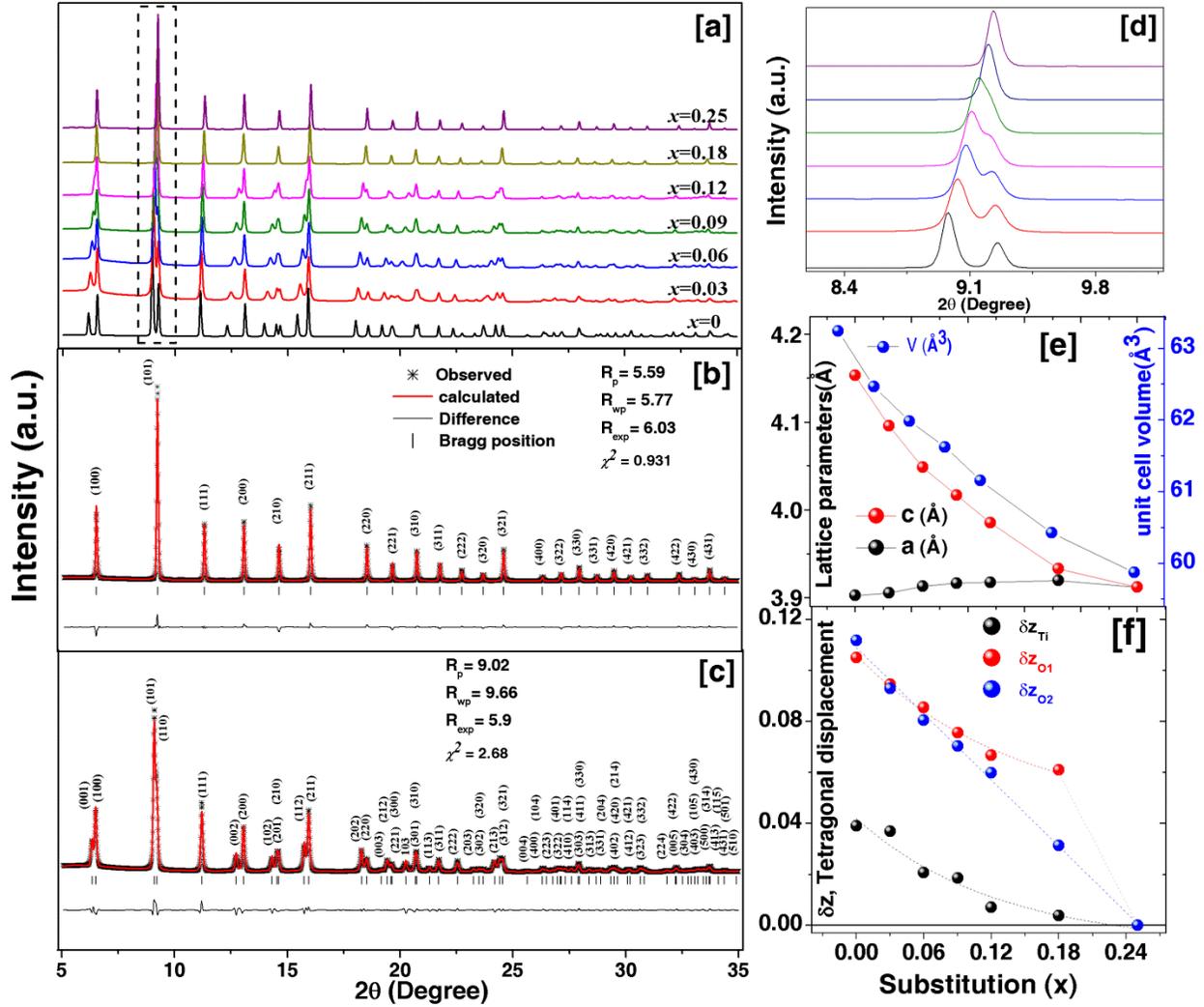

**Figure 1.** (a) Synchrotron based x-ray powder diffraction patterns for $Pb_{(1-x)}La_xTi_{(1-x)}Al_xO_3$ $0 \leq x \leq 0.25$ compositions, (b) Goodness of fitting for $x=0.25$ whereas (c) $x=0.09$ composition, (d) Magnified view of XRD position of (101) and (110) highlighted in Fig.1a, (e) Lattice parameters and volume calculated with the Rietveld refinement from XRD patterns, and (f) Tetragonal displacement of Ti, apical, and planar oxygen position with composition.

Rietveld refinement was performed of synchrotron-based x-ray powder diffraction of PLTA data using Fullprof software [25]. There were no secondary phases observed in the PLTA samples. $PbTiO_3$ tetragonal space group ($P4mm$) was fitted for $0 \leq x \leq 0.18$ compositions whereas, for $x=0.25$ composition was cubic space group ($Pm3m$). After background estimation, Bragg peaks were modelled with Pseudo-Voigt function, using axial divergence asymmetry function. Significant improvement in goodness of fit was obtained with shape and asymmetry parameters refinement. Scale factor, half width parameters ($U, V, W$), lattice parameters ($a, c$) and positional coordinates ($x, y, z$) of various lattice sites were also refined. Starting from pure $PbTiO_3$, XRD data for increasing substitutions were refined one by one using the final refined structural parameters of the lower substituted composition as starting model for the next higher substituted composition. This was done to achieve a trend and a logical structural correlation between samples [26]. Observed and calculated patterns show an excellent match with low values of goodness of fit obtained for all samples. Representative plot of refined XRD data for *x*=0.09 and *x*=0.25 are shown in fig. 1(b-c) and for other samples are provided in Supplementary information (Fig. S1). The fitting R-factors are in acceptable range for all the samples. The variation of calculated lattice parameters with increase in dopant concentration (*x*) is shown in Fig. 1e. Lattice parameter '*a*' slightly increases while '*c*' decreases with increase in substitution. Variations in lattice parameters are observed because of smaller ionic radii of substituted elements compare to parent elements. Also, decrease of volume with the increase of substitution hints of substitution of *Pb/Ti* by *La/Al*. The refined position parameters were determined with Pb ions kept stationary, and refining the positions of the B-site and planar and apical O-ions. Average tetragonal displacement of the B-site from centrosymmetry (*Ti/Al* position obtained from z-values of refinement) is found to decrease with substitution, implying a decrease in tetragonal strain [Fig.1f]. Similar observations are made for the O-ions too.

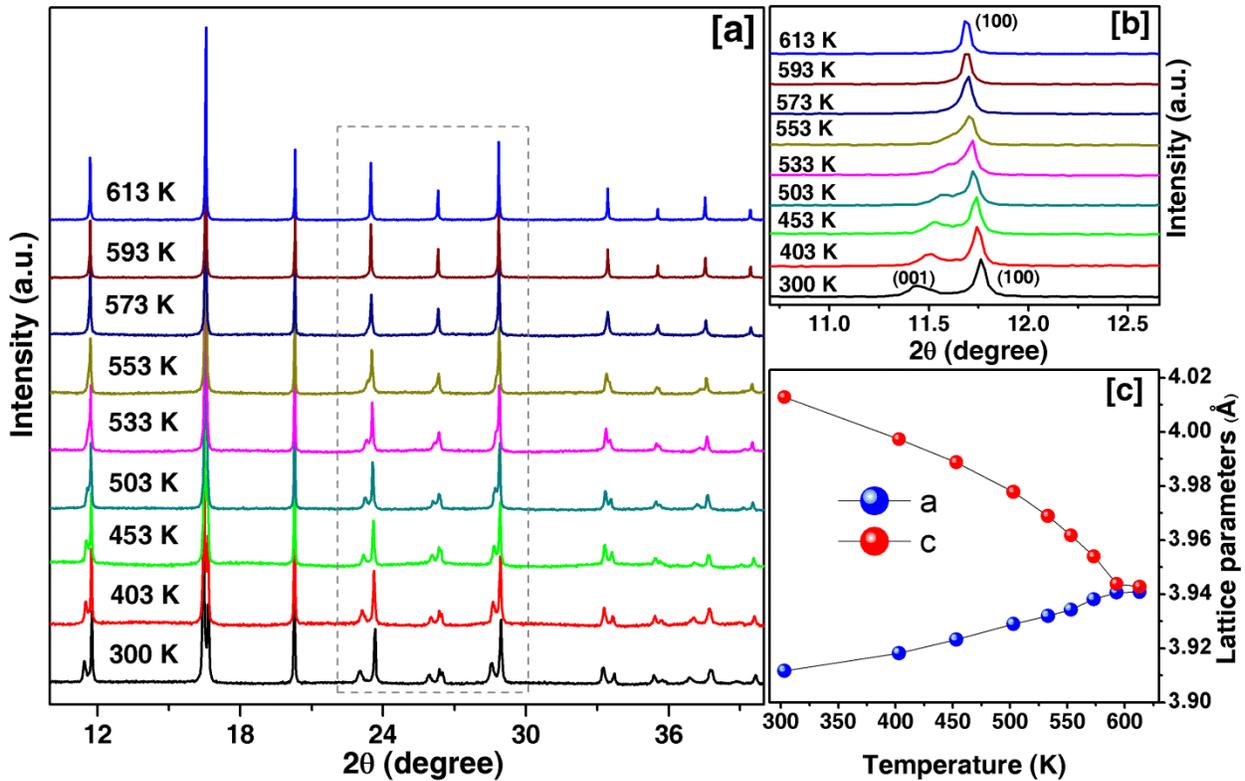

**Figure 2**: Synchrotron-based temperature dependent XRD for *x*=0.09 composition where (a) X-ray spectra at different temperatures (b) merging of (100) and (001) peaks with rising temperature implying tetragonal to cubic phase transition (c) Temperature dependence of lattice parameters.

Temperature-dependent XRD study for $x$=0.09 composition [Fig.2a] shows a large splitting of two peaks (001) and (100) at lower temperatures, indicating a tetragonal phase. With increasing temperature, the peaks merge gradually into a single peak at 573 K, i.e. transform to a cubic phase (Fig. 2b) [27-29]. Likewise, other peaks related to tetragonal to cubic phase show a similar trend. Temperature variation of lattice parameters was calculated by the refinement of temperature-dependent XRD data. A profile matching technique (using Fullprof software) was adopted to calculate the lattice parameters. As shown in Fig.2c, lattice parameter 'a' increases gradually while 'c' decreases considerably with an increase of temperature. Lattice mismatch was calculated using Goldschmidt tolerance factor, $t$ [30].

$$t = \frac{(R_A + R_O)}{\sqrt{2}(R_B + R_O)} \qquad (1)$$

Where $R_A$ and $R_B$ are the average ionic radii of $A$ and $B$-site ions, respectively, while $R_O$, that of oxygen ion. Average ionic radii of ions $Pb^{2+}(XII)$, $La^{3+}(XII)$, $Ti^{4+}(VI)$, $Al^{3+}(VI)$, $O^{2-}(VI)$ are 1.490, 1.360, 0.605, 0.535, and 1.40 Å respectively. The stable perovskite phase will form if $t$ is close to one [29, 31-33]. Untilted perovskites have 0.985 < $t$ < 1.06 range. Tolerance factor will decrease with the decrease of $A$-site ion size. For 0.964 < $t$ < 0.985 range, a tilted anti-phase is expected while t < 0.964 in-phase and anti-phase tilting may occur. If $t$ continues decreases, the stability of perovskite phase will decrease, and eventually, perovskite structure will not form [34]. Tolerance factor for PLTA samples was calculated and found to decrease linearly from 1.019 for $x$=0 to 1.016 for $x$=0.25 composition. Note that the tolerance factor of PLTA samples indicates an untilted stable perovskite structure.

Paraelectric and ferroelectric phases are stabilized by the short and long-range forces, respectively. There is a delicate balance between these two types of forces, and the ferroelectric phase transition occurs as a result of this balance. Long-range Coulomb interaction is affected by defects and domain structure and leads to splitting of longitudinal (*LO*) and transverse (*TO*) optical phonons. Near the phase transition temperature, the short-range and long-range forces compensate each other in such a way that the frequency of optical phonon approaches zero. [35, 36] Hence, Raman spectroscopy is a sensitive and excellent tool in qualitatively assessing retention of domain structure, defect, and structural distortion, thereby understand deformations and lattice strains associated with the substitution in ferroelectric like PLTA samples.

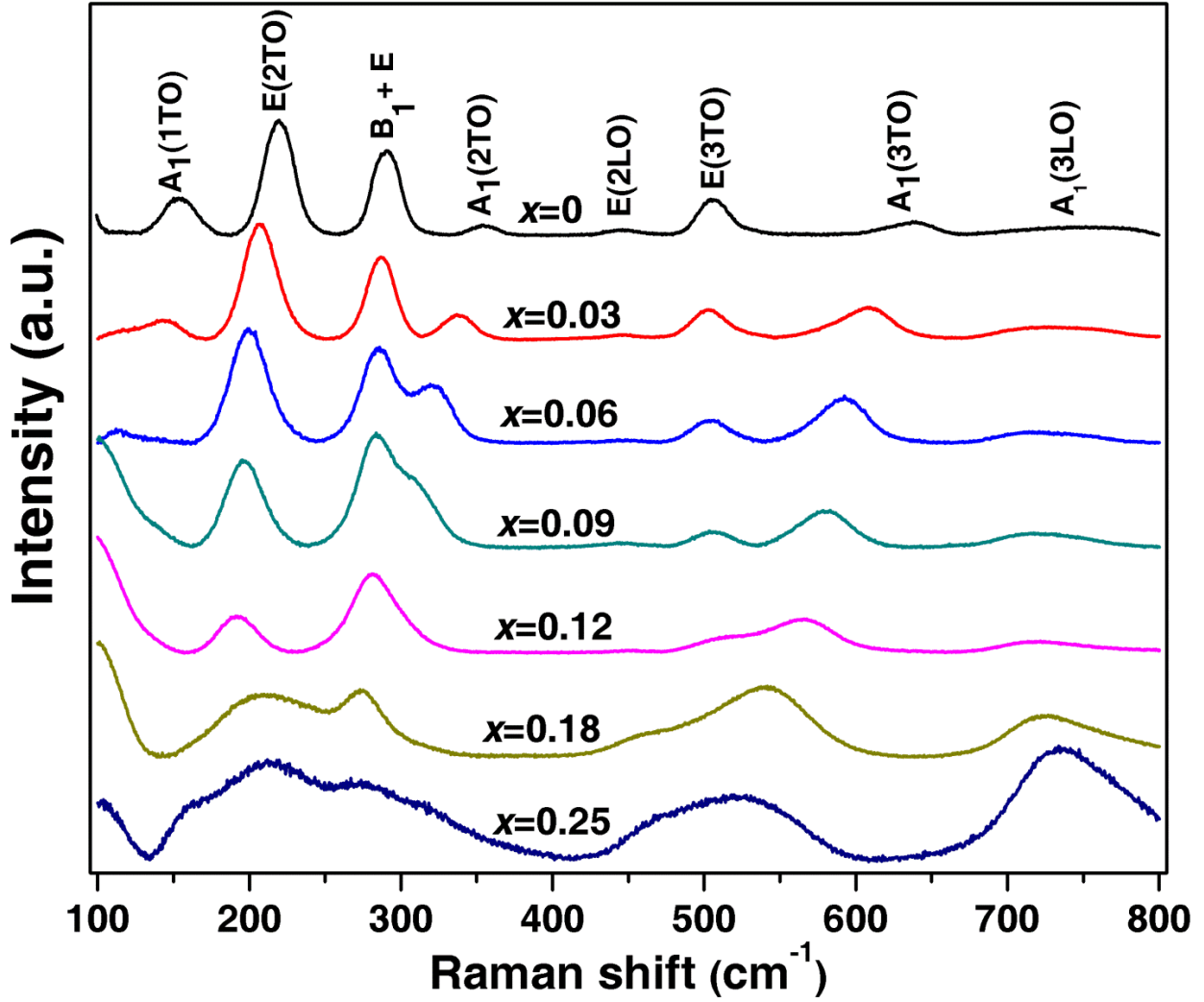

**Figure 3.** Room temperature Raman spectroscopy sprectra of $Pb_{(1-x)}La_xTi_{(1-x)}Al_xO_3$ (0≤x≤0.25) compositions.

Room temperature Raman spectroscopy was performed for all PLTA samples. Raman spectra of PLTA samples are shown in Fig. 3. All the observed modes belong to a pure $PbTiO_3$ structure. For pure $PbTiO_3$ the ferroelectric tetragonal phase ($C_{4v}^1$ (*P4mm*) space group) transform to paraelectric cubic phase ($O_h^1$ (*Pm3m*) space group) at 763 K. At the Γ point in the Brillouin zone, the 12 optical phonon modes transform as the $3T_{1u}+T_{2u}$ irreducible representation. The $T_{1u}$ modes were infrared active, and the $T_{2u}$ mode was silent, neither infrared nor Raman active. Below phase transition temperature, tetragonal phase is called ferroelectric phase and corresponds to $C_{4v}^1$ (*P4mm*) space group. The $T_{1u}$ mode becomes the $A_1+E$ mode, whereas the $T_{2u}$ mode transforms to the $B_1+E$ mode. All these modes were active in both Raman and infrared region. As in the cubic phase, long-range electrostatic forces split the $A_1+E$ modes into *TO* (transverse optical) and *LO* (longitudinal optical) components. Splitting of the $B_1+E$ modes was also allowed, but this had not been observed. $B_1+E$ modes were designated as "silent" modes. The triply degenerate modes of $E(TO)$ (denoted as $E(1TO)$, $E(2TO)$, and $E(3TO)$), along with $3E(LO)$, $3A_1(TO)$, and $3A_1(LO)$ modes are the prominent detectable modes of $PbTiO_3$ [37, 38]. Transverse optical modes of $A_1$ symmetry are important for $PbTiO_3$ based materials because $TO$ modes are oscillations along the $c$-axis, i.e. along the direction of spontaneous polarization. $A_1(1TO)$ mode is due to vibrations of the $B$-site octahedra relative to $A$-site cage. $A_1(2TO)$ and $A_1(3TO)$ soft modes consist of displacements along the c-axis of the $B$-site ion relative to the oxygen and $A$-site ions. It was observed that $A_1(1TO)$, $A_1(2TO)$, and $A_1(3TO)$ modes soften with increasing substitution [Fig. 3] due to tetragonal to cubic structural transformation. Such was also observed in $A/B$- site doped $PbTiO_3$ systems like $Pb_{(1-x)}(Na_{0.5}Sm_{0.5})_xTiO_3$, $PbTi_{1-x}Fe_xO_3$, and $PbZr_{1-x}Ti_xO_3$ [8, 39-41] where $c/a$ ratio reduced. In pure $PbTiO_3$, tetragonal strain originates from hybridization between $Pb(6s^2)$-$O(2p)$ and $O(2p)$-$Ti(3d)$ hybridization [4]. While substituting $Pb/Ti$ by $La/Al$, hybridization becomes softer as $La$ and $Al$ do not hybridize strongly with $O$ atoms. Due to this decrease of hybridization strength, the system tends to transform from tetragonal to a cubic structure. Contrary to other modes, $A_1(3LO)$ mode hardens with substitution. A relative vibration between $Ti$ and planar oxygen ion along $a$ and $b$ axes, this mode becomes more energetic due to an increase in $a$ and $b$ axes with the increase in substitution. Hence, XRD and Raman analyses confirm that the tetragonal strain is relieved with increase in $La$ and $Al$ incorporation.

Field effect scanning electron micrographs of PLTA samples sintered at $1150°C$ for 6h are shown in Fig. 4. Grains are spherical type in shape and are compactly packed in all samples. The average grain size was calculated using Image J software, and it was found to decrease from $12.39\pm3.70$ μm for $x=0.03$ to $0.67\pm0.27$ μm for $x=0.25$ composition. The decrease of grain size could be endorsed to lower diffusivity of rare earth element (here $La$ element) during sintering. Analogous results are also observed in other rare earth doped perovskite and layered perovskite materials[42-44]. Also, the decrease of average grain size could be related to a reduction of oxygen vacancy contribution. Oxygen vacancies are produced in $Pb$ based ceramics by the charge compensation due to the evaporation of volatile $Pb^{2+}$ during the sintering process. Because, the sintering involved the material flow, and in dense ceramics, ionic vacancies are the available sites through which ions can move. Hence oxygen vacancies are the first ionic sites to flow the materials in the dense ceramics.[45, 46] Bulk densities of sintered PLTA pellets were measured by the Archimedes method using double distilled water (density = 1 $g/cm^3$) as immersing liquid. Theoretical densities were also estimated using formula weight and volume of unit cells calculated from the refinement of XRD. Relative density (= Measured density/Theoretical density) was calculated to be ~ 0.92, 0.93, 0.94, 0.95, 0.95, and 0.94 for $x=0.03$, 0.06, 0.09, 0.12, 0.18, and 0.25, respectively.

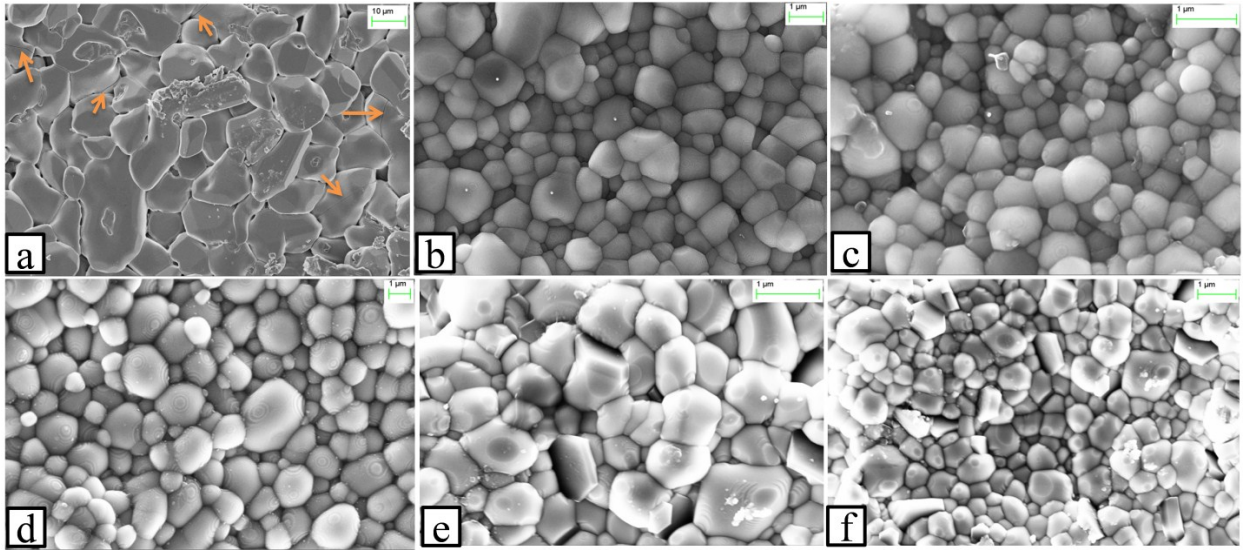

**Figure 4.** FESEM microstructure analysis of $Pb_{(1-x)}La_xTi_{(1-x)}Al_xO_3$ samples where (a) $x$=0.03, (b) $x$=0.06, (c) $x$=0.09, (d) $x$=0.12, (e) $x$=0.18, and (f) $x$=0.25 compositions.

For $x$=0.03, intra-grain cracks were observed (Fig. 4a), highlighted with orange arrows. These fractures occurred due to high tetragonal lattice strain. Different from other samples, the $x$=0.03 has a high tetragonal strain as shown in Fig. 1f. Due to limitations in the cooling process of the furnace, an abrupt anisotropic phase transition from cubic (T $>T_m$) to tetragonal (T$<T_m$) generates cracks, which degrade the ferroelectric properties of the sample. In other substituted samples such inter-grain cracks are not observed, possibly due to lowered tetragonal strain.
1010

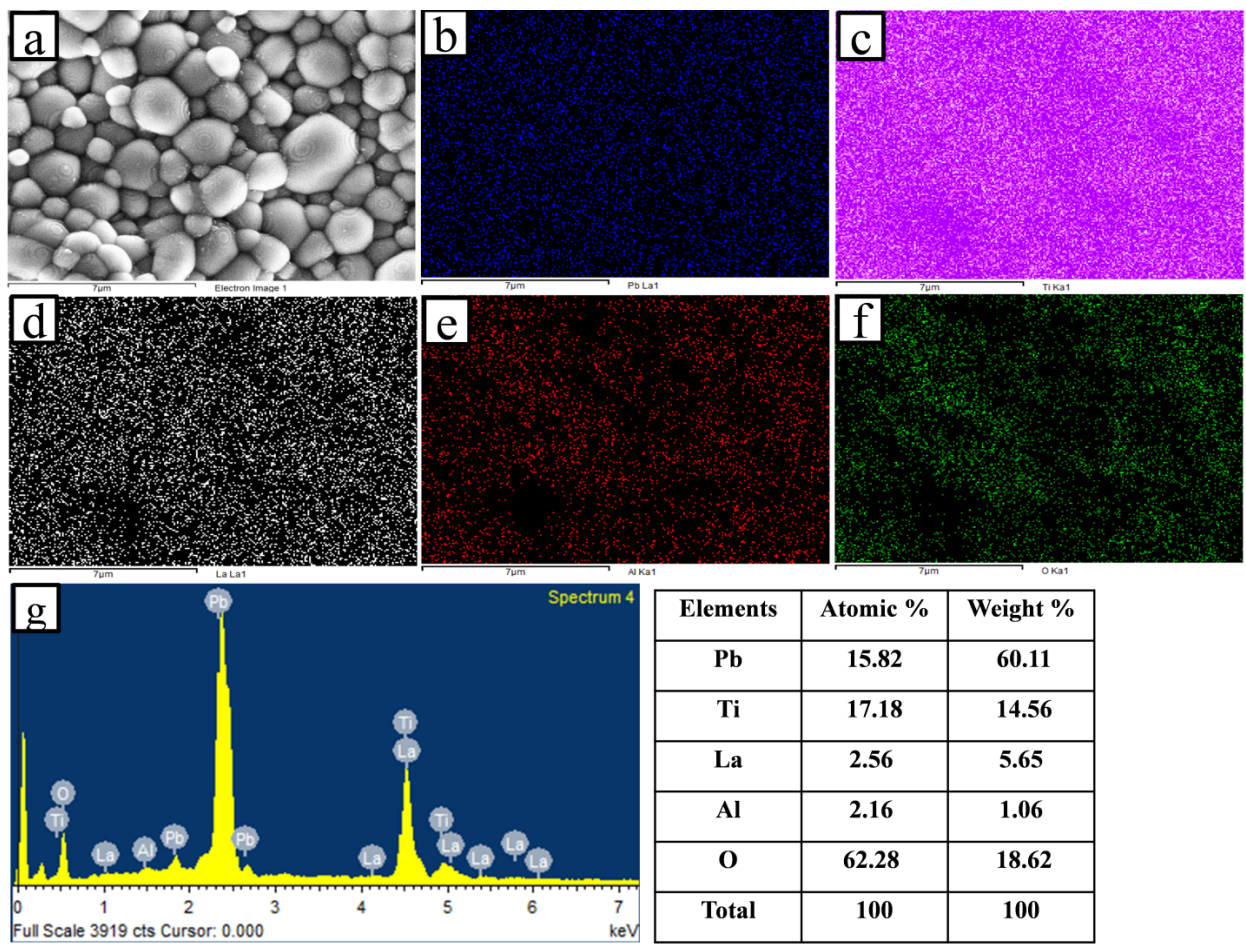

**Figure 5.** (a) SEM-EDS elemental mapping of $Pb_{(1-x)}La_xTi_{(1-x)}Al_xO_3$ for $x$=0.125 composition where (a) secondary electron image and analogous elemental mapping of the element (b) Pb, (c) Ti, (d) La, (e) Al, (f) O elements, (g) Area EDS spectrum and right table for the atomic and weight percentage of various elements.

Elemental mapping by energy dispersive x-ray spectrometry (SEM/EDS) was performed for all the samples. Obtained compositions from EDS analysis of individual samples were compared to the corresponding targeted chemical phase. The obtained compositions matched well within the typical errors related to EDS analysis. A representative EDS spectrum, and elemental mapping along with the secondary electron micrograph for the $x$=0.125 is shown in Fig. 5(a-g). Spot EDS was carried out on different grains to confirm chemical compositions. Elemental analysis for area EDS was performed to verify the homogeneity of the samples [Fig. 5(b-f)]. An integrated area spectrum is shown in Fig. 5g and tabulated with atomic and the weight percentage of constituent elements.

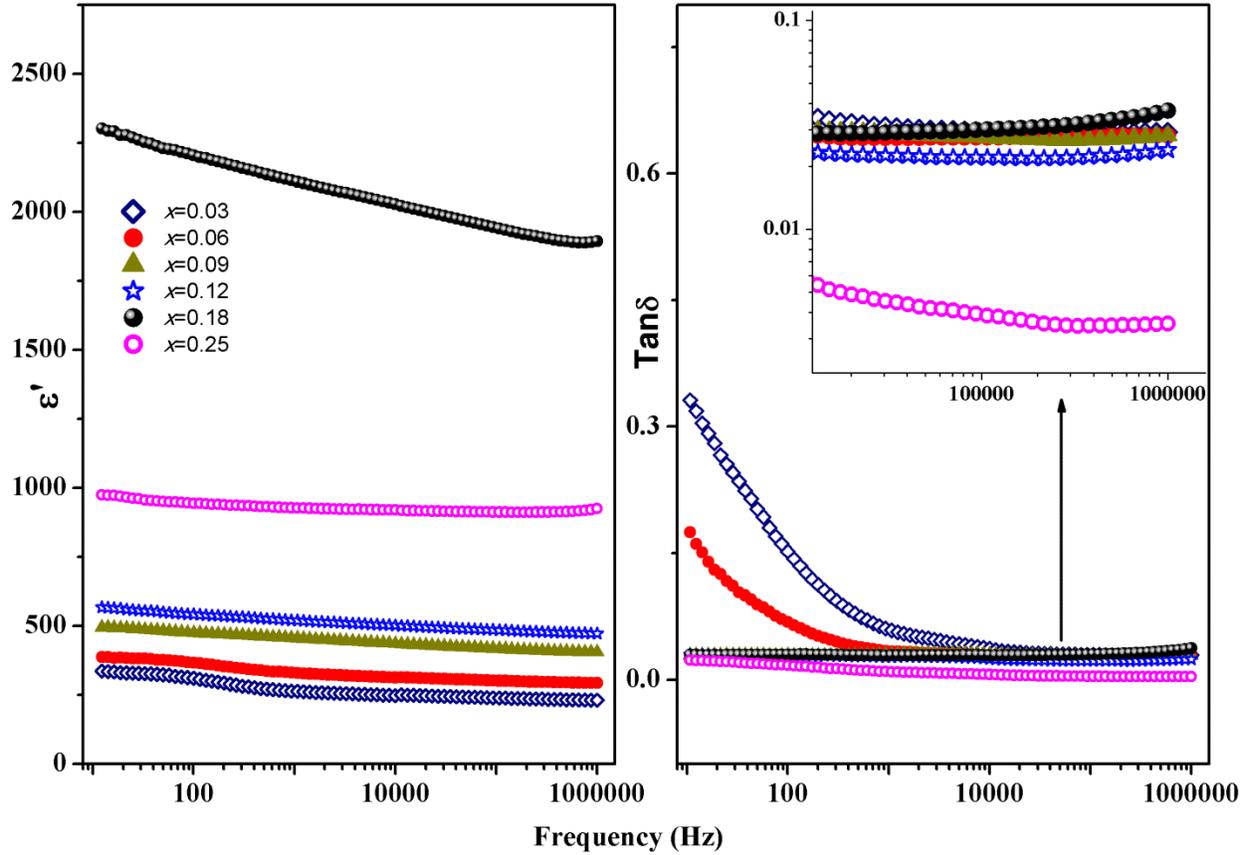

**Figure 6**. (a) Room temperature dielectric constant versus frequency for $Pb_{(1-x)}La_xTi_{(1-x)}Al_xO_3$ samples, and (b) Room temperature tanδ versus frequency for $Pb_{(1-x)}La_xTi_{(1-x)}Al_xO_3$ samples.

**Dielectric Properties**

Frequency dependent dielectric properties were measured at room temperature for samples with compositions, $x$=0.03 to 0.25 and are shown in Fig. 6a. As discussed in introduction section it is difficult to fabricate pure $PbTiO_3$ ($x$=0) ceramic pellet, hence, its dielectric and ferroelectric properties are not investigated in the present work. Dielectric constant ($\varepsilon'$) and loss (tanδ) are slightly decreasing with increase in frequency for all samples [Fig. 6]. This is in agreement with similar reports on other perovskite titanates.[42, 47, 48] Room temperature $\varepsilon'$ increases notably while *tanδ* (Fig. 6b) slightly increases with increasing $x$ in the range 0.03≤$x$≤0.18. This can be correlated with higher electronic polarizability of *La* than *Pb*[49]. Also, decease of the phase transition temperature with the increase in composition may be reason of high dielectric constant with increase in $x$. A very high dielectric constant and low loss (tanδ) for the $x$=0.18 composition, making the material usable for the capacitor applications etc. However, for $x$=0.25, $\varepsilon'$ suddenly reduces. As paraelectric samples have low dielectric constant than ferroelectric materials [42], this decrease may be due to the paraelectric behaviour of $x$=0.25. Note that, except $x$=0.25, all other samples are in ferroelectric phase at room temperature, as revealed from XRD and Raman spectroscopic results.

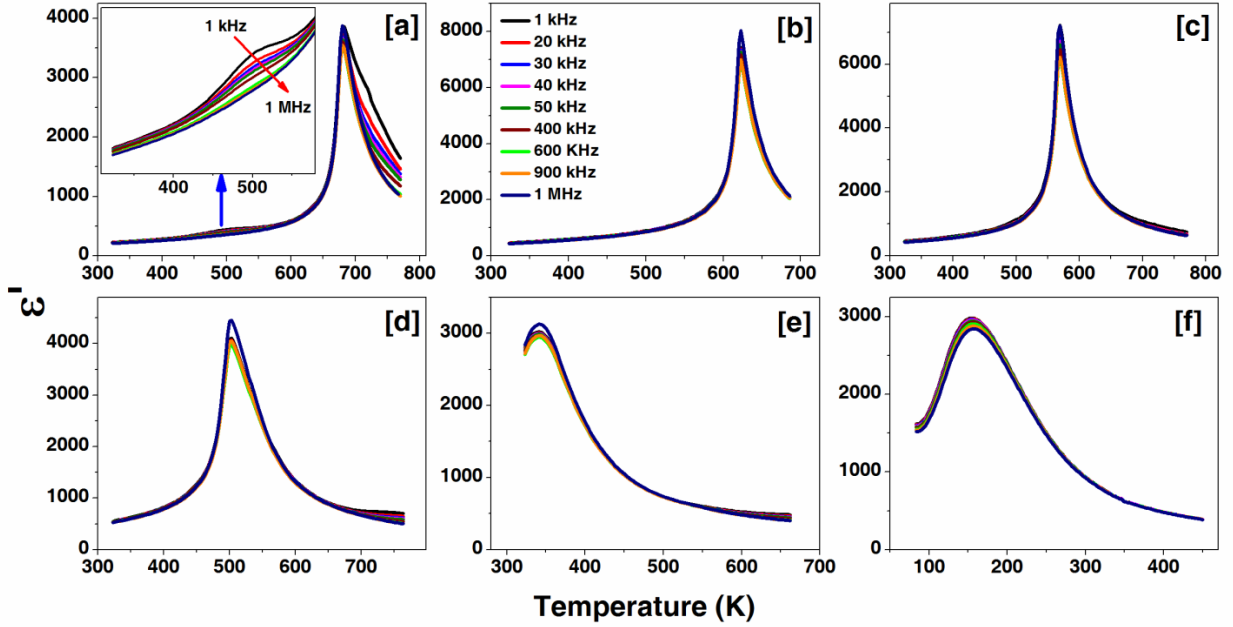

**Figure 7.** Temperature-dependent dielectric constant at different frequencies showing phase transition for $Pb_{(1-x)}La_xTi_{(1-x)}Al_xO_3$ samples where (a) $x=0.03$, (Inset figure is showing an anomalies ~ 500 K at low frequencies whereas at high frequency, there is no change) (b) $x=0.06$, (c) $x=0.09$, (d) $x=0.12$, (e) $x=0.18$, and (f) $x=0.25$ composition.

Temperature-dependent $\varepsilon'$ and $tan\delta$, frequencies in 1 kHz – 1 MHz range, were measured in the temperature range, 300-773 K, to study phase transition temperature of samples with $0.03 \leq x \leq 0.18$ [Fig. 7(a-e)]. For $x=0.25$, same was measured in the range 80-450 K [Fig. 7f]. Temperature-dependent $tan\delta$ curves are shown in Fig. S2. A ferroelectric to paraelectric phase transition is generally associated with a $\varepsilon'$ maximum at $T_m$, due to dielectric catastrophic behaviour. Pure $PbTiO_3$ undergoes a sharp ferroelectric to paraelectric phase transition around 763 K (Curie temperature) accompanied by a structural transition from polar tetragonal to non-polar cubic phase [50]. Also, same tetragonal to cubic phase transition was obtained for pure $PbTiO_3$ at 11.2 GPa at room temperature [51]. In PLTA samples, $T_m$ linearly decreases gradually from 680 K (for $x=0.03$) to ~175 K (for $x=0.25$) with an increase of *La* and *Al* substitutions. For $0.03 \leq x \leq 0.18$ composition, the $\varepsilon'$ maxima are frequency independent. For $x=0.03$, a dielectric anomaly around 500 K is observed in $\varepsilon'$-$T$ curve, for lower frequencies [Fig. 7a inset]. At higher frequencies this is absent. This anomaly is related to relaxation process involving oxygen vacancies.[52, 53] Note that with increased substitution this relaxation is not observed. This is direct evidence that oxygen defects reduce with increasing *x* values.

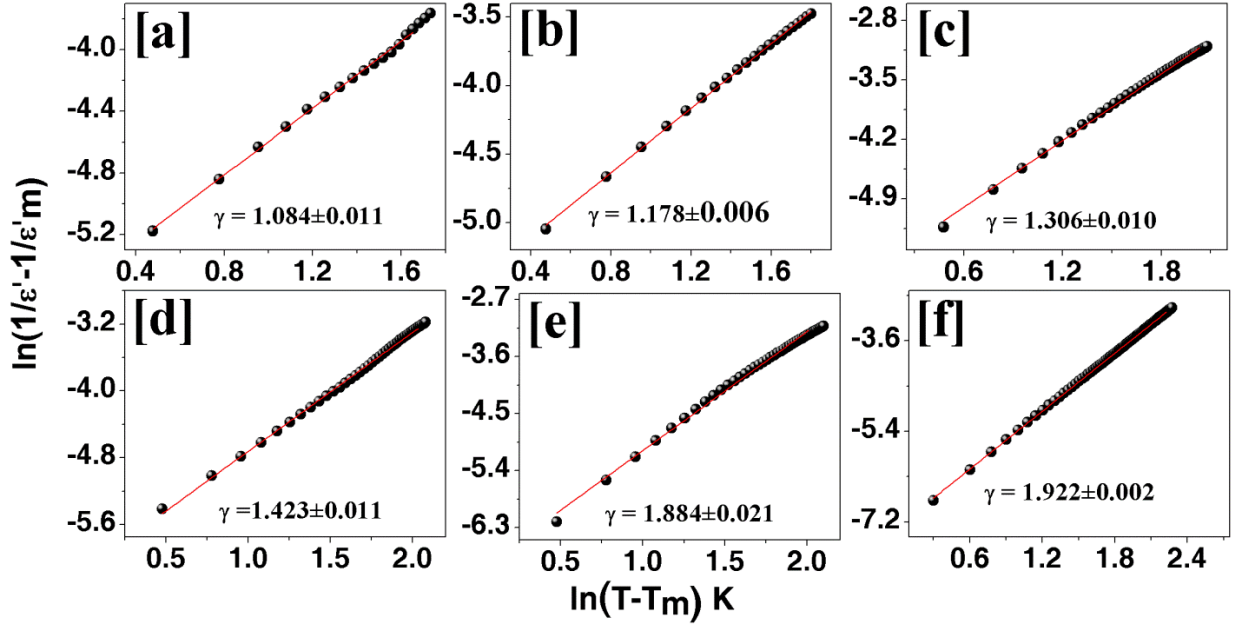

**Figure 8**. Modified Curie Weiss law was modelled for the PLTA samples of ln1/ε'-1/ε'$_m$ verus ln(T-T$_m$)K curves (a) *x*=0.03, (b) *x*=0.06, (c) *x*=0.09, (d) *x*=0.12, (e) *x*=0.18, and (f) *x*=0.25 compositions. Symbols are representing the experimental data whereas red line fitted data.

The diffuseness of *ε′-T* maxima was observed for all PLTA samples. A modified Curie-Weiss model describes the diffused phase transition[54, 55],

$$\frac{1}{\varepsilon'} - \frac{1}{\varepsilon'_m} = C^{-1}(T - T_m)^\gamma \qquad (2)$$

where, *C* is Curie-Weiss constant and *γ* (1≤*γ*≤2) gives the degree of diffuseness. A sharp transition is achieved for *γ*=1, while *γ*=2 is an ideal diffuse phase transition. The degree of diffuseness was calculated by the least square linear fitting of ln ($\frac{1}{\varepsilon'}$ - $\frac{1}{\varepsilon'_m}$) versus ln (T-T$_m$) curves at a frequency of 1 MHz of all the PLTA ceramic samples. Linear fitted data are shown in Fig. 8. The constant *γ*, obtained from the slope of linear fits was found to increase with the increase in La/Al incorporation. The *γ* was found from 1.084±0.011 to 1.922±0.002 for *x*=0.03 to 0.25 composition. This increase of diffuseness indicates compositional disorder originating from the random distribution of *La* and *Al*. Broadening of the maxima instead of sharp transition relates to nature of structural changes at Curie point. This is usually observed in perovskites with a random distribution of ions on structurally identical sites in lattice.[56, 57]

Hence, from XRD, Raman and dielectric properties, with increasing *La* and *Al* incorporation, it can be inferred that tetragonal strain is relieved on account of the random distribution of *La* and *Al*, thereby decreasing *T$_m$*, and affecting relaxation properties.

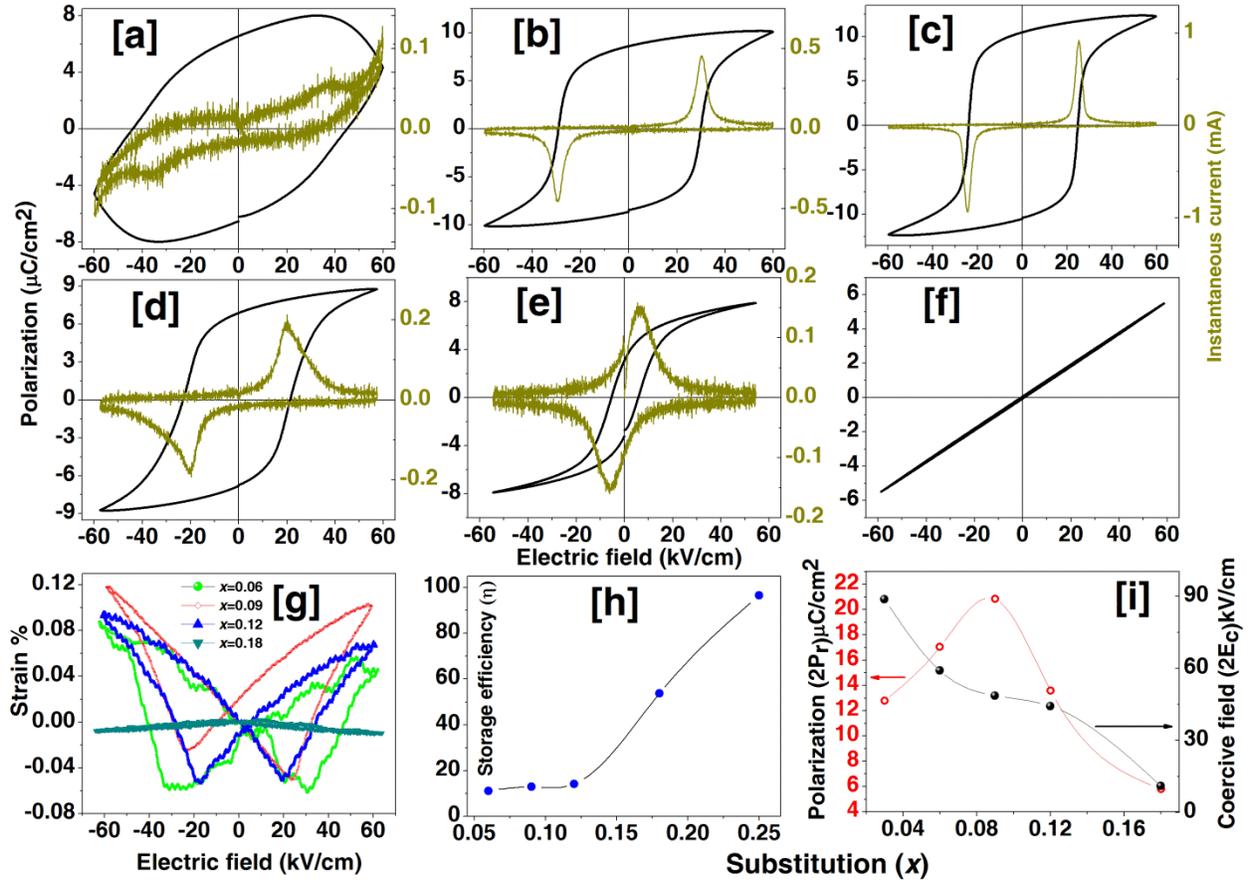

**Figure 9.** Polarization and instantaneous current versus electric field at room temperature of Pb$_{(1-x)}$La$_x$Ti$_{(1-x)}$Al$_x$O$_3$ samples where (a) $x$=0.03, (b) $x$=0.06, (c) $x$=0.09, (d) $x$=0.12, (e) $x$=0.18, (f) $x$=0.25 compositions, (g) Bipolar strain versus electric field for $x$=0.06, 0.09, 0.12, and 0.18 composition, (h) Storage efficiency versus composition, and (i) Spontaneous polarization and coercive field versus composition.

**Ferroelectric properties**

A strongly nonlinear hysteretic polarization as a function of applied *ac* electric field (E), at room temperature, was observed for 0.03≤$x$≤0.18 composition. The hysteresis loops were measured at a frequency of 1 Hz. Proper ferroelectric saturated loops are obtained for 0.06≤$x$≤0.18 compositions as shown in Fig 9(a-e). This is most probably due to switching of domains under the influence of *E* [58]. Hysteretic properties ensure spontaneous polarization of these materials. Spontaneous polarization and coercive field versus compositions are shown in Fig. 9i. For $x$=0.09 composition, spontaneous polarization is maximum (2P$_r$ = 20.841 μm/cm$^2$) whereas coercive field is low (2E$_c$ = 48.708 kV/cm). Instantaneous current versus electric field showed peaks related to domain switching behaviour in these compounds. These peaks confirmed proper ferroelectric nature of these compounds. For $E < E_c$, (critical electric field required to switch the ferroelectric domains), a linear nature of polarization is observed instead of a hysteretic type [59]. However, for $x$=0.25, absolute linear polarization was observed suggesting paraelectric nature. This is in agreement with XRD, Raman and dielectric analyses. There was no signature of domain switching for $x$=0.25 composition.

We observe that domain switching is not pronounced in $x$=0.03 composition. In high $Pb$ containing materials, there is a serious problem of $Pb$ loss due to its extreme volatile nature during sintering at elevated temperatures. It is widely known that the ionization of oxygen vacancy will create [60].

$$O_O = V_O + \frac{1}{2}O_2 \qquad (3)$$

Conduction of electrons created by the ionization of oxygen vacancies could be described using the Kröger–Vink notation of defects[61, 62].

$$V_O = V_O^{\cdot} + e' \qquad (4)$$

$$V_O^{\cdot} = V_O^{\cdot\cdot} + e' \qquad (5)$$

where $V_O^{\cdot}, V_O^{\cdot\cdot}$ were single-ionized and doubly-ionized oxygen vacancies respectively. The ionization of oxygen vacancies will create electrons. These electrons might be loosely bonded to the $Pb^{2+}$ or $Ti^{4+}$ ions, and their exact location depends on many factors[63, 64] The overall reactions could be described as:

$$Pb_{Pb}^{\times} + O_O^{\times} \rightarrow PbO (\uparrow) + V_{Pb}'' + V_O^{\cdot\cdot} \qquad (6)$$

Oxygen vacancies formed to maintain the charge neutrality in PLTA ceramics are known as extrinsic vacancies [60]. These oxygen vacancies can hop easily due to their high mobility in the applied high electric field and accumulate in the places with low free energy, such as domain walls and interfaces with electrodes. Accumulating of these oxygen vacancies at the domain boundary causes domain pinning which restricts polarization switching.[61, 65] Due to larger proportions of $Pb$ in $x$=0.03, O-vacancies are more probable, large free charges restricting to apply electric field more than coercive field.

Bipolar strain versus electric-field curves of PLTA ceramics measured at room temperature under an applied maximum electric field 60 kV/cm is shown in [Fig. 9g]. Typical butterfly-shaped loops, for 0.06≤$x$≤0.12, confirm piezoelectric nature of the samples. A piezoelectric strain was found to decrease with substitution. Similar behaviour is also observed in other perovskites.[66, 67] For $x$>0.18, samples did not exhibit piezoelectric nature.

Energy storage density for a polarizable material is given by, 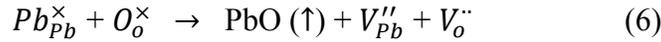 $W = \int_{P_i}^{P_f} E dP$, where, $W$ represents the electrical energy density, $E$ refers to the applied external electric field, $P_i$ and $P_f$ are the initial and final polarization respectively. [68-70] In a ferroelectric material, the charging and discharging process corresponds to the total energy supplied, $W_{supplied}$, and stored, $W_{stored}$, respectively. These are actually the areas under the $E$-$P$ diagrams for charging and discharging plots. The area enclosed by the ferroelectric loop, therefore, corresponds to the energy lost, $W_{lost}$. A relation between these energy densities is given by, $W_{supplied} = W_{stored} + W_{lost}$. The storage

efficiency, η, can be estimated by, $\eta = \frac{W_{stored}}{W_{supplied}}$. For PLTA samples, an area of loops gradually decreases with increasing substitution. Hence, the energy loss is less. Therefore, the storage efficiency, η, increases with *La/Al* addition [Fig. 9h]. As a result, ferroelectric materials in the paraelectric region can have high storage efficiency. In case of *x*=0.25, this was calculated to be 96%, and can be used as an energy storage material. On the other hand, for 0.06≤*x*≤0.18) samples are in ferroelectric phase with nonlinear hysteresis curve at room temperature. Thereby, losses are more and cannot be used as energy storage materials, but can be used in ferroelectric RAM, sensors etc. We have plotted the composition dependence of energy storage density and observe that for the initial doping of *x*≤0.12 the efficiency is less and rising nominally, whereas, for *x*≥0.18 it rises drastically.

**Conclusions**

$Pb_{(1-x)}La_xTi_{(1-x)}Al_xO_3$ (0≤*x*≤0.25) ceramics were fabricated form the powders synthesize successfully using the sol-gel process. Reitveld refinement of synchrotron XRD data was performed for all samples to investigate the effect of La/Al doping on the crystal structure. Analysis confirmed the tetragonal *P4mm* phase for the 0≤*x*≤0.18 composition whereas *x*=0.25 composition was found to be in cubic structure with *Pm3m* space group. Lattice parameters '*c*' decreased while '*a*' increased with the increase in substitution (*x*). Raman spectroscopy shows variations in intensity and energy of phonons modes related to tetragonal to cubic phase transformation. A temperature-dependent dielectric study confirmed the phase transition from tetragonal to the cubic structure. The decrease in the temperature of phase transition with increased substitution was correlated with the XRD and Raman spectroscopy. Temperature-dependent x-ray diffraction study co-relates the exact structural phase transition from tetragonal to cubic structure. The degree of diffuseness of the dielectric peak was found to increases with increase in substitution. We have obtained good ferroelectric/piezoelectric loops for the samples with 0.06≤*x*≤0.18. Energy storage density was calculated for all the samples and sample with composition *x*=0.25 showed the highest energy density of 96.5%.

**Acknowledgements**
One of the authors is thankful to University Grants Commission for providing research fellowship (NFO-2015-17-OBC-UTT-28455). Principle investigator expresses sincere thanks to Indian Institute of Technology Indore for funding the research. The authors sincerely thank Sophisticated Instrument Centre (IIT Indore) for FESEM studies, Mr. Ravindra Jangir and Mr. Ashok Bhakar (RRCAT, Indore) for low-temperature dielectric measurements and Mr. Rituraj Sharma (IISER Bhopal) for Raman studies. One of the authors (Dr. Sajal Biring) acknowledges the financial support from Ministry of Science and Technology, Taiwan (MOST 105-2218-E-131-003 and 106-2221-E-131-027).

**Supplementary Information**

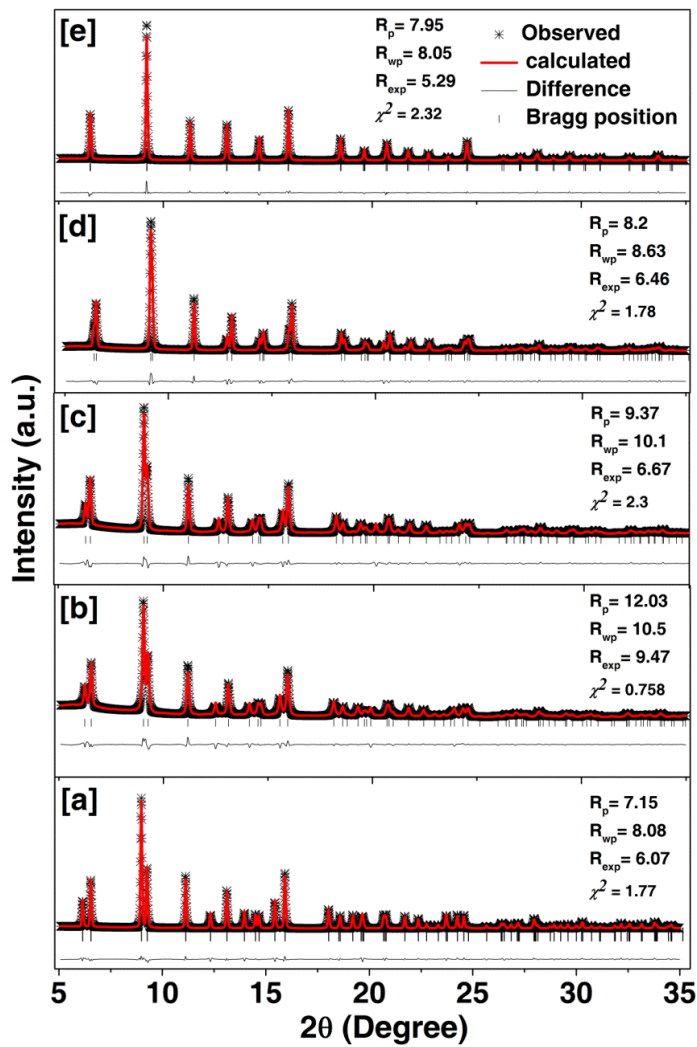

**Figure S1**. Goodness of fitting after Rietveld refinement are shown for the PLTA samples where (a) $x=0$, (b) $x=0.03$, (c) $x=0.06$, (d) $x=0.12$, and (e) $x=0.18$ compositions.

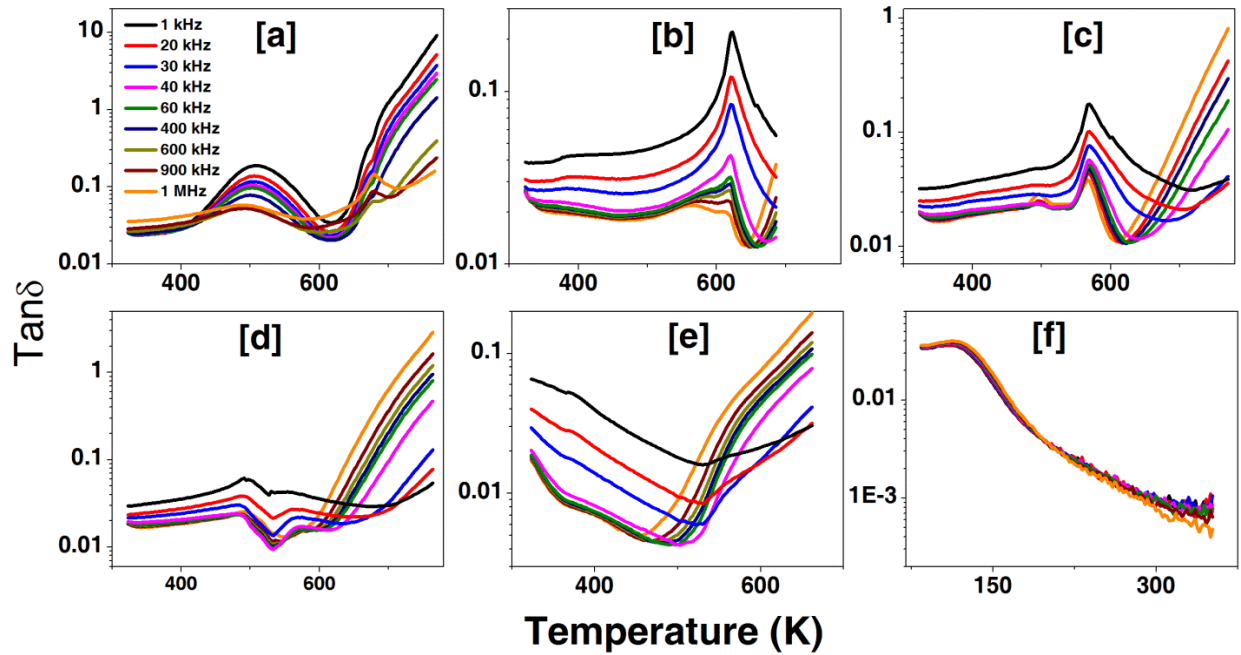

**Figure S2.** Temperature-dependent Tanδ of the PLTA samples at different frequencies where (a) $x$=0.03, (b) $x$=0.06, (c) $x$=0.09, (d) $x$=0.12, (e) $x$=0.18, and (f) $x$=0.25 compositions.